\documentclass[twocolappendix,11pt,a4paper]{emulateapj}

\usepackage{psfrag}
\usepackage{psfig}
\usepackage{pspicture}
\usepackage{graphicx}
\usepackage[export]{adjustbox}
\usepackage{amssymb,amsmath}
\usepackage{natbib}

\def\apj{ApJ}

\submitted{published in the \apj}

\begin{document}

\title {Spectroscopic Confirmation of a Coma Cluster Progenitor at \lowercase{$z\sim$} 2.2}

\author{
Behnam Darvish\altaffilmark{1}, 
Nick Z. Scoville\altaffilmark{1},
Christopher Martin\altaffilmark{1},
David Sobral\altaffilmark{2},
Bahram Mobasher\altaffilmark{3},
Alessandro Rettura\altaffilmark{4},
Jorryt Matthee\altaffilmark{5},
Peter Capak\altaffilmark{6},
Nima Chartab\altaffilmark{3},
Shoubaneh Hemmati\altaffilmark{6},
Daniel Masters\altaffilmark{6},
Hooshang Nayyeri\altaffilmark{7},
Donal O'Sullivan\altaffilmark{1},
Ana Paulino-Afonso\altaffilmark{8},
Zahra Sattari\altaffilmark{3},
Abtin Shahidi\altaffilmark{3},
Mara Salvato\altaffilmark{8},
Brian C. Lemaux\altaffilmark{9},
Olivier Le F{\`e}vre\altaffilmark{10},
Olga Cucciati\altaffilmark{11}
}

\setcounter{footnote}{0}
\altaffiltext{1}{Cahill Center for Astrophysics, California Institute of Technology, 1216 East California Boulevard, Pasadena, CA 91125, USA; email: bdarv@caltech.edu; bdarv001@ucr.edu}
\altaffiltext{2}{Department of Physics, Lancaster University, Lancaster, LA1 4YB, UK}
\altaffiltext{3}{Department of Physics and Astronomy, University of California, Riverside, 900 University Avenue, Riverside, CA 92521, USA}
\altaffiltext{4}{W.M. Keck Observatory, 65-1120 Mamalahoa Highway, Waimea, HI 96743, USA}
\altaffiltext{5}{Department of Physics, ETH Z{\"u}rich, Wolfgang-Pauli-Strasse 27, CH-8093 Z{\"u}rich, Switzerland}
\altaffiltext{6}{IPAC, Mail Code 314-6, California Institute of Technology, 1200 East California Boulevard, Pasadena, CA 91125, USA}
\altaffiltext{7}{Department of Physics and Astronomy, University of California Irvine, Irvine, CA 92697, USA}
\altaffiltext{8}{Max-Planck-Institut f{\"u}r extraterrestrische Physik (MPE), Giessenbachstrasse 1, D-85748 Garching bei M{\"u}nchen, Germany}
\altaffiltext{9}{Department of Physics, University of California, Davis, One Shields Avenue, Davis, CA 95616, USA}
\altaffiltext{10}{Aix Marseille Universit{\'e}, CNRS, LAM (Laboratoire d'Astrophysique de Marseille), UMR 7326, F-13388 Marseille, France}
\altaffiltext{11}{INAF - Osservatorio di Astrofisica e Scienza dello Spazio di Bologna, via Gobetti 93/3, I-40129 Bologna, Italy}

\begin{abstract}

We report the spectroscopic confirmation of a new protocluster in the COSMOS field at $z$ $\sim$ 2.2, COSMOS Cluster 2.2 (CC2.2), originally identified as an overdensity of narrowband selected H$\alpha$ emitting candidates. With only two masks of Keck/MOSFIRE near-IR spectroscopy in both $H$ ($\sim$ 1.47-1.81 $\mu$m) and $K$ ($\sim$ 1.92-2.40 $\mu$m) bands ($\sim$ 1.5 hr each), we confirm 35 unique protocluster members with at least two emission lines detected with S/N $>$ 3. Combined with 12 extra members from the zCOSMOS-deep spectroscopic survey (47 in total), we estimate a mean redshift and a line-of-sight velocity dispersion of $z_{mean}$=2.23224 $\pm$ 0.00101 and $\sigma_{los}$=645 $\pm$ 69 km s$^{-1}$ for this protocluster, respectively. Assuming virialization and spherical symmetry for the system, we estimate a total mass of $M_{vir}$ $\sim$ $(1-2) \times$10$^{14}$ $M_{\odot}$ for the structure. We evaluate a number density enhancement of $\delta_{g}$ $\sim$ 7 for this system and we argue that the structure is likely not fully virialized at $z$ $\sim$ 2.2. However, in a spherical collapse model, $\delta_{g}$ is expected to grow to a linear matter enhancement of $\sim$ 1.9 by $z$=0, exceeding the collapse threshold of 1.69, and leading to a fully collapsed and virialized Coma-type structure with a total mass of $M_{dyn}$($z$=0) $\sim$ 9.2$\times$10$^{14}$ $M_{\odot}$ by now. This observationally efficient confirmation suggests that large narrowband emission-line galaxy surveys, when combined with ancillary photometric data, can be used to effectively trace the large-scale structure and protoclusters at a time when they are mostly dominated by star-forming galaxies.\\

\textit{Unified Astronomy Thesaurus concepts:} Galaxy clusters (584); High-redshift galaxy clusters (2007); High-redshift
galaxies (734); Large-scale structure of the universe (902); Galaxy evolution (594); Star formation (1569); Galaxy
environments (2029);

\end{abstract}


\section{Introduction} \label{int}

Galaxy clusters and protoclusters at high redshifts ($z\gtrsim$ 2) are ideal laboratories for studying structure formation, cosmology, and the effect of early environments on galaxy formation and evolution. The latter is particularly important as the $z\sim$ 2-3 redshift regime traces the peak of star-formation and AGN activity in the universe \citep{Madau14,Khostovan15}, when many physical processes, such as cold gas flow into galaxies, outflow and feedback processes, mergers, and likely environment governed the evolution of galaxies. 

At low redshift, the relation between galaxy properties and environment is relatively well established. However, at high redshifts ($z\gtrsim$ 2), there are conflicting results, partly due to the small number of confirmed structures, and often having only a small number of confirmed members. 

\begin{figure*}
\centering
\includegraphics[width=7.0in]{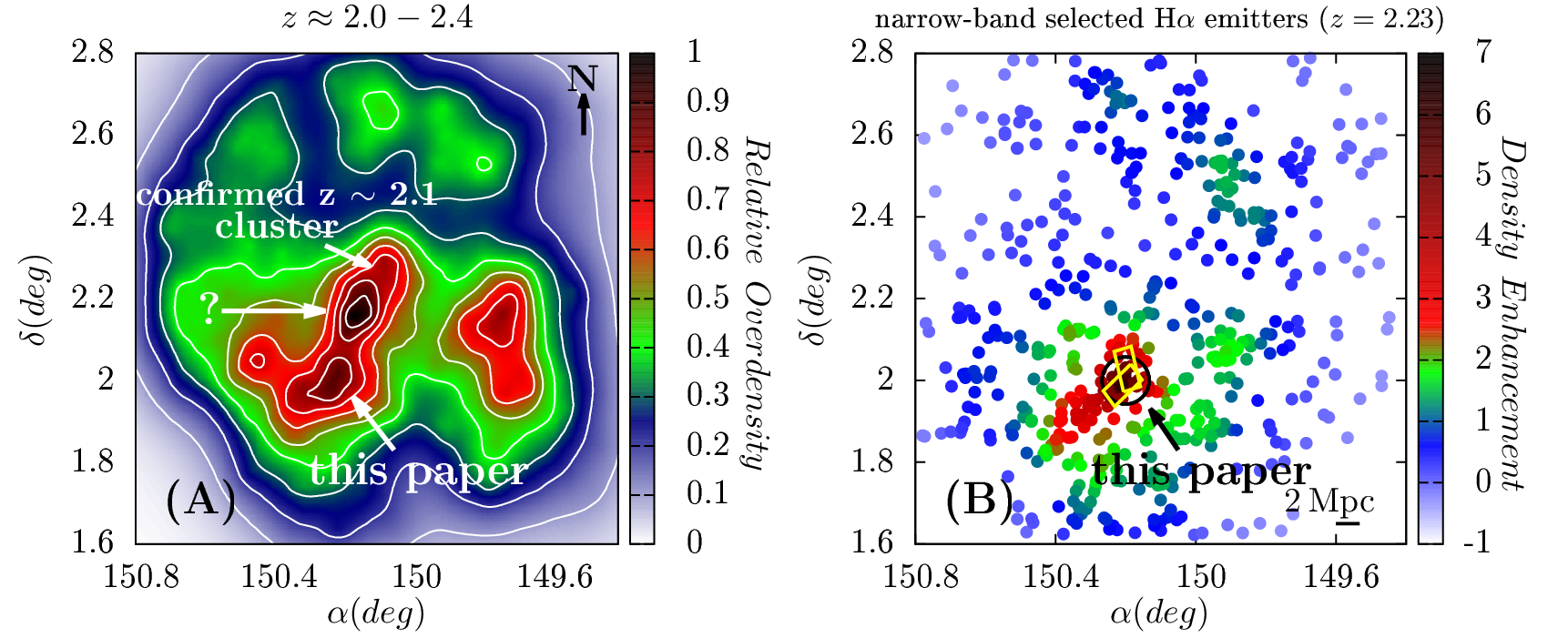}
\caption{(A) Relative overdensity map in the COSMOS field for a redshift slice centered at $z$=2.23 (redshift width of $\approx$ $\pm$ 0.2). The map is adaptively smoothed using a weighted adaptive Gaussian kernel \citep{Darvish15a,Darvish17} with a global kernel width of 2 Mpc. An extended, several megaparsec-scale LSS is clearly seen. The northern section of this LSS is a spectroscopically confirmed cluster at $z$ $\sim$ 2.1 \citep{Yuan14}. There is evidence for some overdensity in the middle section of this structure (shown with a question mark). There is also an extended southern section to this structure. (B) Spatial distribution of narrowband selected H$\alpha$ emitter candidates from the HiZELS/COSMOS survey \citep{Sobral13} at $z$ $\sim$ 2.23 (redshift width of $\sim$ 0.03-0.04) color-coded by their density enhancement. The southern section of the extended LSS (left panel) is clearly seen as an overdensity of narrowband selected H$\alpha$ emitting candidates. We perform follow-up spectroscopic observations targeting the densest region of this southern section shown with a black circle of 2 Mpc radius. The positions of the spectroscopic masks (Section \ref{Strategy}) are shown with yellow rectangles. Note the $z$ $\sim$ 2.1 cluster and the potential central overdensity (shown with the question mark on the left panel) are not seen here given the narrowness of the narrowband filter.}
\label{fig:map}
\end{figure*}

At $z\gtrsim$ 2, there is poor agreement between current studies on the mass-metallicity relation, with results varying from an absence of any environmental trends \citep{Kacprzak15}, to an enhancement \citep{Shimakawa15} or a deficiency of metals \citep{Valentino15} for star-forming galaxies in denser environments. The situation is the same regarding the relation between environment and star-formation activity in galaxies at $z\gtrsim$ 2 (e.g.; see \citealp{Darvish16,Shimakawa18,Chartab20}) and the environmental dependence of the gas content of galaxies (e.g.; see \citealp{Lee17,Noble17,Darvish18b,Hayashi18,Wang18a,Tadaki19}). The discrepant results are likely caused by different dynamical states of the environments probed, different selection functions, small sample sizes, AGN contamination, different star formation rate (SFR), metallicity, and gas mass indicators used, complications due to extinction correction, and so on. This implies the need for finding more high-$z$ structures with a well-defined sample of galaxies and a large number of confirmed spectroscopic measurements.

Cluster candidates at high redshifts can be detected through the concentration of quiescent galaxies (e.g.;  \citealp{Strazzullo15}), by probing the environment of highly rare and active systems, such as quasars, radio and submillimeter galaxies, and Ly$\alpha$ blobs (e.g.; \citealp{Matsuda04,Venemans07,Capak11,Wylezalek13}), or an overdensity of IR sources with, e.g. {\it Spitzer}, {\it Herschel}, or {\it Planck} (e.g.; \citealp{Papovich10,Muzzin13b,Clements14,Rettura14}). These approaches have led to the spectroscopic confirmation of a number of candidate structures at $z$ $\gtrsim$ 2 (e.g.; \citealp{Capak11,Cucciati14,Lemaux14,Yuan14,Wang16,Cucciati18}; also see the review by \citealp{Overzier16}). The detection and spectroscopic confirmation of clusters traced by passive galaxies is hard because of the small number density of quiescent galaxies at higher reshifts and the lack of emission lines in their spectra which makes the spectroscopic observations challenging. Moreover, the rarity of very active galaxies such as quasars in the present-day surveys makes the high-$z$ protocluster detection probed by them difficult.

An observationally efficient and physically motivated technique to identify protoclusters at $z$ $\gtrsim$ 2 is to target concentrations of emission-line galaxies, such as H$\alpha$ and Ly$\alpha$ emitters using narrowband filters (e.g.; \citealp{Matsuda11,Koyama13b}). The high concentration of star-forming, emission-line systems (prior to quenching) in protoclusters has been theoretically predicted by the hierarchical galaxy formation models and has successfully resulted in the spectroscopic confirmation of some protoclusters and large-scale structures (LSSs) at $z$ $\gtrsim$ 2 (e.g.; \citealp{Chiang15,Lemaux18}).Therefore, large emission-line galaxy surveys can be used to effectively trace the LSSs and protoclusters at $z$ $\gtrsim$ 2. 

Here, we report the spectroscopic confirmation of a protocluster, dubbed COSMOS Cluster 2.2 (CC2.2), originally found as an overdensity of narrowband selected H$\alpha$ emitters at $z\sim$ 2.2 in the High-Z Emission Line Survey (HiZELS;  \citealp{Geach12,Sobral13,Sobral14}) of the Cosmic Evolution Survey (COSMOS) field \citep{Scoville07}. In Section \ref{motivation}, we explain the protocluster selection. In Section \ref{observation}, we present the spectroscopic observations and equip them with ancillary spectroscopic data. The protocluster properties and its fate are presented in Section \ref{property}. The results are compared with other high-$z$ protoclusters in Section \ref{comparison}. We summarize the results in Section \ref{sum}.

Throughout this paper, we assume a flat $\Lambda$CDM cosmology with $H_{0}$=70 km s$^{-1}$ Mpc$^{-1}$, $\Omega_{m}$=0.3, and $\Omega_{\Lambda}$=0.7. Unless otherwise stated, the transverse cosmological distances are presented as physical distances. The ``physical'' scale at the redshift of the protocluster ($z$ $\sim$ 2.23) is $\sim$ 0.5 Mpc per arcmin.

\begin{figure*}
\centering
\includegraphics[width=7.0in]{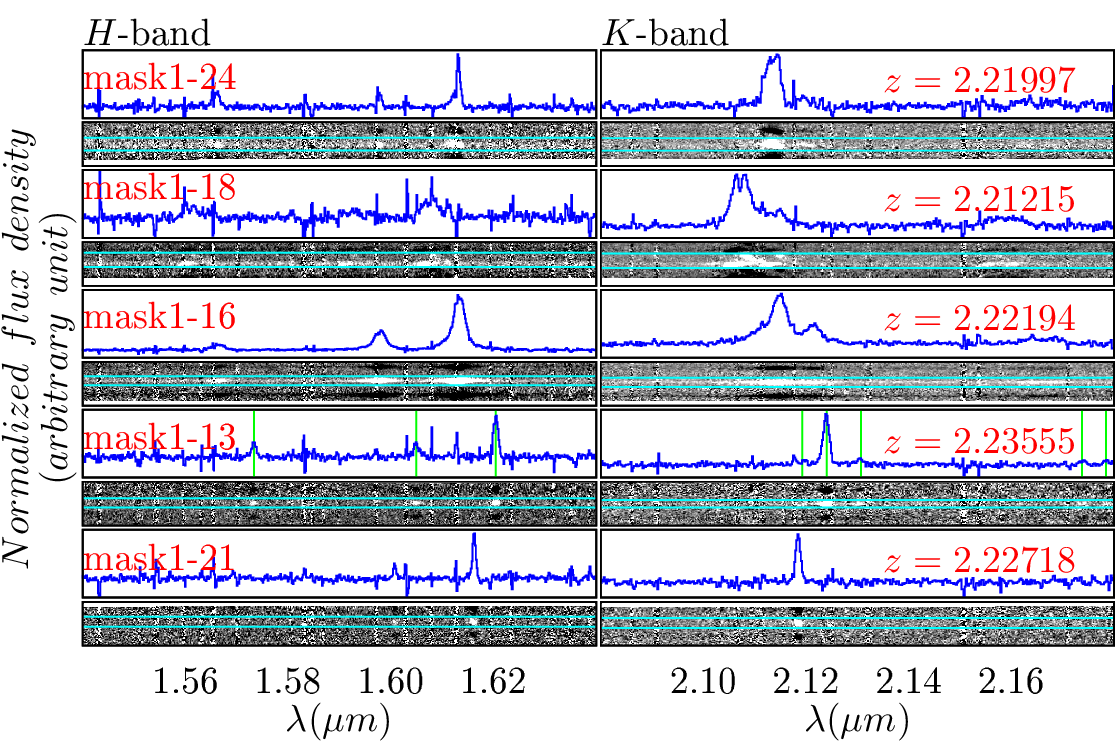}
\caption{Example 2D and extracted 1D spectra showing some emission lines. Cyan lines show the 1D extraction window. The position of H$\beta$, [O{\sc iii}]$\lambda$4959, [O{\sc iii}]$\lambda$5007, [N{\sc ii}]$\lambda$6549, H$\alpha$, [N{\sc ii}]$\lambda$6583, [S{\sc ii}]$\lambda$6717, and [S{\sc ii}]$\lambda$6731 emission lines is shown with vertical green lines for one of the galaxies. The top two spectra show two merger cases, the third one is a broad-line AGN, and the last two spectra  show normal star-forming galaxies in the protocluster.}
\label{fig:spectra}
\end{figure*}         

\section{Protocluster Selection} \label{motivation}

Figure \ref{fig:map} (A) shows the relative overdensity map in the COSMOS field for a redshift slice centered at $z$=2.23, with a width $\pm$1.5$\sigma_{\Delta z/(1+z)}$ $\approx$ $\pm$ 0.2 from the center of the slice \citep{Darvish17}, where $\sigma_{\Delta z/(1+z)}$ is the typical photometric redshift uncertainty at $z$ $\sim$ 2.2 \citep{Laigle16}. In making this map, all galaxies more massive than the mass completeness limit ($\geqslant$ 10$^{10}$ $M_{\odot}$) at this redshift are selected. In addition, $\geqslant$ 90\% of the photometric redshift probability distribution function of these galaxies should lie within the boundaries of this redshift slice. The map is adaptively smoothed using a weighted adaptive Gaussian kernel \citep{Darvish15a} with a global kernel width of 2 Mpc. An extended, several megaparsec-scale LSS is clearly seen. The northern section of this LSS is a spectroscopically confirmed cluster at $z$ $\sim$ 2.1 \citep{Yuan14}. There is evidence for some conspicuous overdensity in the middle section of this structure. There is also an extended southern section to this structure, which is the focus of this paper. 

Figure \ref{fig:map} (B) clearly reveals this southern section. Here, we show the spatial distribution of uniformly probed narrowband selected H$\alpha$ emitter candidates from the HiZELS survey \citep{Sobral13} in the COSMOS field at $z$ $\sim$ 2.23 (Section \ref{sample}). They are color-coded by their density enhancement defined as $\frac{\Sigma-\Sigma_{0}}{\Sigma_{0}}$, where $\Sigma$ is the surface number density and $\Sigma_{0}$ is the mean surface number density. This southern section stands out as an overdensity of H$\alpha$ emitters (see also \citealp{Geach12}). We perform follow-up spectroscopic observations with Keck/MOSFIRE targeting the densest region of this southern section as a potential protocluster (shown with a black circle in Figure \ref{fig:map} (B)). 

\begin{table*}
\begin{center}
\caption{Coordinate and Redshift of the Sources} 
\begin{scriptsize}
\centering
\begin{tabular}{lccccccc}
\hline
\hline
\noalign{\smallskip}
Number & R.A. & Decl. & Spectroscopic $z$ & ID(HiZELS) & ID(COSMOS)\footnote{COSMOS IDs and $K_{s}$ magnitudes are from the \citealp{Laigle16} catalog.} & $K_{s}$(COSMOS) & Comment\\
& (deg) & (deg) & & & & (mag) &\\
\hline
\\
mask1-1 & 150.184235 & 2.035242 & 2.23227 & S12B-1073 & 483880 & 21.053 & primary\\
mask1-2 & 150.162308 & 1.999728 & 2.23796 & S12B-1133 & 461469 & 21.780 & primary\\
mask1-3 & 150.162231 & 1.997168 & 2.23962 & S12B-1142 & 459801 & 22.225 & primary\\
mask1-4 & 150.178958 & 2.009019 & - & - & 465362 & 16.753 & 2MASS star\\
mask1-5 & 150.197937 & 2.026497 & 2.23767 & S12B-1089 & 478717 & 22.143 & primary\\
mask1-51 & 150.200721 & 2.023885 & 2.23675 & - & 476338 & 23.899 & serendipitous\\
mask1-6* & 150.179947 & 1.992562 & 2.23276 & S12B-1149 & 457031 & 23.414 & primary\\
mask1-7 & 150.201492 & 2.011835 & 2.22076 & S12B-1115 & 469074 & 22.781 & primary\\
mask1-8 & 150.207275 & 2.015360 & 2.22390 & S12B-1110 & 471600 & 21.757 & primary, merger?\\
mask1-81 & 150.208559 & 2.014025 & 2.24683 & - & 470941 & 22.171 & serendipitous\\
mask1-9 & 150.214371 & 2.013219 & 2.23730 & S12B-1111 & 470543 & 21.507 & primary\\
mask1-91* & 150.213577 & 2.014169 & 2.23653 & S12B-1108 & 470634 & 22.683 & serendipitous\\
mask1-10 & 150.208500 & 2.002617 & 2.04398 & - & 463455 & 22.508 & filler, field\\
mask1-11 & 150.215375 & 2.004858 & 2.21980 & - & 464709 & 23.769 & filler\\
mask1-12 & 150.210208 & 1.995008 & 2.99093 & - & - & - & filler, field\\
mask1-13 & 150.208420 & 1.989571 & 2.23555 & S12B-9026 & 455052 & 22.451 & primary\\
mask1-14 & 150.209686 & 1.983837 & 2.23588 & S12B-9096 & 451484 & 22.204 & primary, merger\\
mask1-15 & 150.216417 & 1.988758 & 2.22850 & - & 455603 & 20.372 & filler\\
mask1-16 & 150.230774 & 1.998720 & 2.22194 & S12B-1139 & 462238 & 21.365 & primary\\
mask1-17 & 150.218689 & 1.981452 & 2.23929 & S12B-9145 & 450967 & 22.385 & primary\\
mask1-18 & 150.217958 & 1.969297 & 2.21215 & - & 443583 & 20.477 & filler, triple merger?\\
mask1-19 & 150.235428 & 1.984851 & 2.22909 & S12B-9103 & 452250 & 22.433 & primary\\
mask1-20 & 150.227844 & 1.954604 & 2.22964 & S12B-9563 & 433445 & 22.825 & primary\\
mask1-21 & 150.252487 & 1.980309 & 2.22718 & S12B-9161 & 449433 & 23.109 & primary\\
mask1-22 & 150.242096 & 1.963336 & 2.24096 & S12B-9425 & 439360 & 22.871 & primary\\
mask1-23 & 150.247604 & 1.963640 & 2.24408 & S12B-9419 & 439051 & 23.362 & primary\\
mask1-24 & 150.262009 & 1.974060 & 2.21997 & S12B-9256 & 446347 & 21.155 & primary, merger\\ 
mask2-1 & 150.226868 & 2.069255 & - & S12B-3033 & 506226 & 22.186 & primary\\
mask2-2 & 150.183044 & 2.077852 & 2.23340 & S12B-3052 & 511651 & 21.499 & primary\\
mask2-3 & 150.224152 & 2.055650 & 2.22964 & S12B-1036 & 496918 & 21.971 & primary\\
mask2-4 & 150.214042 & 2.046742 & - & - & 491994 & 22.396 & filler\\
mask2-5 & 150.178958 & 2.009019 & - & - & 465362 & 16.753 & 2MASS star\\ 
mask2-6 & 150.194167 & 2.038033 & 2.11461 & - & 485225 & 23.099 & filler, field\\
mask2-7 & 150.199921 & 2.031268 & 2.23029 & S12B-1080 & 481208 & 23.040 & primary\\
mask2-8 & 150.214966 & 2.021282 & 2.23626 & S12B-1097 & 475366 & 21.447 & primary, merger\\
mask2-9 & 150.213974 & 2.019010 & 2.23866 & S12B-1105 & 473829 & 22.526 & primary, merger\\
mask2-10* & 150.213577 & 2.014169 & 2.23653 & S12B-1108 & 470634 & 22.683 & primary\\
mask2-11 & 150.214042 & 2.046742 & - & - & 467174 & 23.073 & filler\\
mask2-12 & 150.209702 & 1.990308 & 2.23784 & S12B-9015 & 455204 & 23.440 & primary\\
mask2-13 & 150.163498 & 2.000660 & 2.23259 & S12B-1138 & 461703 & 23.766 & primary\\
mask2-14* & 150.179947 & 1.992562 & 2.23243 & S12B-1149 & 457031 & 23.414 & primary\\
mask2-15 & 150.207657 & 1.981512 & 2.22801 & S12B-9144 & 450160 & 22.865 & primary, merger?\\
mask2-16 & 150.199478 & 1.979534 & 2.23906 & S12B-9175 & 449105 & 22.216 & primary\\
mask2-17 & 150.168716 & 1.985868 & 2.23499 & S12B-9080 & 454336 & 20.690 & primary\\
mask2-18 & 150.171083 & 1.982756 & 2.23571 & - & 451149 & 21.578 & filler\\
\\
\hline
\\
zDEEP-404985 & 150.129107 & 1.990073 & 2.2252 & - & 455565 & 22.460 & ancillary\\            
zDEEP-426887 & 150.130297 & 2.009929 & 2.2371 & - & 467708 & 23.504 & ancillary\\            
zDEEP-404921 & 150.134173 & 1.985729 & 2.2412 & S12B-9081 & 452539 & 23.006 & ancillary\\            
zDEEP-427277 & 150.141604 & 2.046844 & 2.2328 & S12B-1053 & 490796 & 23.362 & ancillary\\            
zDEEP-405266 & 150.146074 & 2.006951 & 2.2351 & S12B-1120 & 465895 & 23.224 & ancillary\\            
zDEEP-404838 & 150.161394 & 1.981538 & 2.2311 & - & 450533 & 22.327 & ancillary\\            
zDEEP-418470 & 150.164187 & 1.982856 & 2.2239 & - & 451562 & 22.958 & ancillary\\            
zDEEP-426933 & 150.166506 & 2.014942 & 2.2340 & - & 471112 & 23.597 & ancillary\\            
zDEEP-427537 & 150.178728 & 2.069249 & 2.2269 & S12B-3032 & 505282 & 24.327 & ancillary\\            
zDEEP-426643 & 150.209795 & 1.986637 & 2.2310 & S12B-9070 & 453251 & 23.086 & ancillary\\            
zDEEP-405942 & 150.214480 & 2.044393 & 2.2298 & - & 490170 & 22.578 & ancillary\\            
zDEEP-418791 & 150.232958 & 2.025177 & 2.2245 & - & 477167 & 23.677 & ancillary\\
\\
\hline
\hline
\label{table1}
\end{tabular}
\end{scriptsize}
\end{center}
\end{table*}      

\section{Spectroscopic Observations} \label{observation}

\subsection{Sample Selection for Spectroscopy} \label{sample}

To increase the success rate of our spectroscopic observations, we focus on potential targets in the vicinity of the candidate protocluster that are likely emission-line galaxies (e.g.; star-forming, starburst, or AGN). This is because detecting emission lines is easier and observationally more efficient than finding absorption features in the stellar continuum, which require longer integration times. Moreover, the strongest absorption features appear around the rest-frame 4000 \AA \ and are then redshifted to the $J$ band at the presumed redshift of the protocluster, a region populated by many atmospheric absorption and emission features. 

Hence, as the primary targets in the vicinity of the overdensity, we rely on the narrowband selected H${\alpha}$ emitting candidates from the HiZELS survey \citep{Sobral13} in the COSMOS field at $z$ $\sim$ 2.23. These are detected as excess color in the UKIRT/WFCAM and VLT/HAWK-I narrowband $K$ filters (centered at $\lambda$ $\sim$ 2.12-2.13 $\mu$m with an FWHM $\Delta\lambda$ $\sim$ 200-300 \AA , corresponding to a redshift width of $\Delta z$ $\sim$ 0.03-0.04 centered at $z$ $\sim$ 2.23-2.24) relative to the broadband $K$ filter.
To minimize contamination from other emission lines, if available, a combination of double-line detections (in both narrowband $K$ and $H$ and/or $K$ and $J$), broadband color-color selections ($Z-K$ versus $B-Z$ and $B-R$ versus $U-B$), and photometric redshift cuts (1.7 $<$ $z_{phot}$ $<$ 2.8) were also implemented. This primary target list is complete down to an H$\alpha$ flux of $\gtrsim$ 1$\times$10$^{-17}$ erg s$^{-1}$cm$^{-2}$, rest-frame EW(H$\alpha$+[N{\sc ii}]) $\geqslant$ 25 \AA, observed SFR $\gtrsim$ 3 $M_{\odot} yr^{-1}$ (Chabrier initial mass function), and stellar mass limits of $\gtrsim$ 10$^{9.7}$ $M_{\odot}$ (see \citealp{Sobral13,Sobral14} for details).

In our design of the multiobject spectroscopic masks, we also added filler objects. They are selected from either the latest COSMOS2015 $K_{s}$-band selected or the previous $I$-band selected photometric redshift catalogs \citep{Ilbert09,Laigle16}. The fillers are selected to be in the vicinity of the overdensity, classified as star-forming galaxies (to increase their detection rate) based on their rest-frame $NUV-r$ versus $r-J$ colors \citep{Ilbert13}, with their photometric redshift in the range 1.7 $<$ $z_{phot}$ $<$ 2.8. Given their selection, some of the fillers may belong to the potential protocluster as well.        
  
\subsection{Observational Strategy} \label{Strategy}

The observations were conducted on 2018 December 8 and 2019 January 13-15 with KeckI/MOSFIRE NIR multiobject spectrograph under clear conditions with the average seeing of $\sim$ 0.5$^{\prime \prime}$-0.6$^{\prime \prime}$ in December and $\sim$ 0.3$^{\prime \prime}$-0.4$^{\prime \prime}$ in January. Given the expected redshift of the structure, we perform observations in both $K$ ($\sim$ 1.92-2.40 $\mu$m) and $H$ ($\sim$ 1.47-1.81 $\mu$m) bands to cover emission lines that can later be used to measure the SFR (H$\alpha$ or H$\beta$), nebular extinction (H$\beta$ and H$\alpha$), gas-phase metallicity ([N{\sc ii}]$\lambda$6549, [N{\sc ii}]$\lambda$6583, and H$\alpha$), electron density ([S{\sc ii}]$\lambda\lambda$6717,6731 doublet), source of ionization (BPT diagram), and ionization state of the gas ([O{\sc iii}]$\lambda$4959, [O{\sc iii}]$\lambda$5007, and H$\beta$) for galaxies. 

We designed two masks in the vicinity of the protocluster candidate (Figure \ref{fig:map}). They were designed in such a way to maximize the number of primary targets. The masks contained unique sources except for one source that would later be used to estimate systematics. In total, we placed 30 unique primary targets and 9 fillers on the masks. 

A 2MASS star per mask was used to estimate the observing conditions, such as the seeing and the spatial profile of point sources. Using an ABBA dithering pattern, we observed each mask in each filter for a total exposure time of $\sim$ 72-96 minutes with a midpoint airmass of $\sim$ 1.0-1.3. Using sky lines, we estimate an FWHM observed spectral resolution of $\sim$ 4.5 \AA \ and $\sim$ 6 \AA \ in $H$ and $K$ bands, respectively, with the slit width of 0.7$^{\prime \prime}$. These correspond to $R$ $\sim$ 3600 and $\delta z$ $\sim$ 0.0003. 

\subsection{Data Reduction} \label{Reduction} 
            
We used the MOSFIRE DRP to reduce the data. The reduction involves flat-fielding, cosmic-ray removal, sky subtraction, and vacuum wavelength calibration on a slit-by-slit basis. The outputs are the 2D spectra and their uncertainties. We extract the 1D spectrum and its associated error using the optimal extraction algorithm of \cite{Horne86}. This is done by weighted summing of fluxes in an optimized window around the 2D spectrum, where the weights incorporate both the flux uncertainties and the spatial extent of the 2D spectrum (spatial profile). To determine the optimized window, we use the spatial profile of each source. To extract the spatial profile, we collapse the 2D spectrum of each source along the wavelength direction in the vicinity of bright, high signal-to-noise (S/N) features and then fit a Gaussian function to the profile. We choose the optimized window as $\pm$3 $\times$ the standard deviation of the spatial profile around its center. If determining the spatial profile fails because of, e.g., faint, low S/N spectrum, we instead rely on the spatial profile of our 2MASS star. In a few cases (e.g.; nearby merging systems) where determining the optimized window is tricky, we instead extract the 1D spectra in a boxcar window wide enough to fully cover all the features (e.g.; Fig. \ref{fig:spectra} second example). Finally, for all the sources, we visually check the extraction window to make sure that the fluxes are fully measured. Figure \ref{fig:spectra} shows some example 2D and their extracted 1D spectra. 

\subsection{Redshift Estimation} \label{Redshift} 

Table \ref{table1} lists the extracted redshifts for our spectroscopic sample in the two masks, as well as the coordinate, HiZELS ID (for primary sources), $K_{s}$ magnitude, the COSMOS ID of each source (based on a match with the COSMOS2015 catalog with a 1$^{\prime\prime}$ radius), and whether a source is a primary target, a filler, a serendipitous detection, a potential merger, or a field galaxy. We report a secure redshift for galaxies that have at least two significant (S/N $\geqslant$ 3) emission lines. The reported redshift is the average redshift that we obtain based on the peak of all the available emission lines for each source (mostly H$\alpha$ and [O{\sc iii}]$\lambda$5007). For sources that show signs of mergers in their spectra and/or in their images (commented as ``merger'' in Table \ref{table1}), the average redshift of different components is given.

To check for systematics in redshifts for objects on different masks, one object is observed twice (mask1-6* and mask2-14*). The extracted redshift difference for this source is $\sim$ 0.0003, similar to the resolution of $\delta z$ $\sim$ 0.0003. Another primary object is also observed twice, with a serendipitous detection in the other mask (mask2-10* and mask1-91*). The extracted redshift difference for this source is zero. To check for systematics in obtaining redshifts in different bands ($H$ and $K$), we compare redshifts obtained based on emission lines in each individual band (if available). The absolute difference is in the range $\Delta z(HK)$=0.00009-0.00218 with a median value of 0.00029, similar to the redshift resolution of $\delta z$ $\sim$ 0.0003. To further check the reliability of redshifts, for objects whose emission lines can be fitted with a single Gaussian function, we also determine redshift by fitting a Gaussian. In all cases, the extracted redshifts are within $\sim$ 0.0003 of what we originally determined.

Out of 30 unique primary targets (commented as ``primary'' in Table \ref{table1}), 29 yield secure redshifts at $z$ $\sim$ 2.23, showing the robustness of narrowband selection (when combined with further photometric information) in tracing the LSS at high redshift. This also shows that with modest spectroscopic observations ($\sim$ 1-2 hr), true high-$z$ clusters can be efficiently confirmed. We also find some fillers and serendipitous detections with spectroscopic redshifts in the vicinity of the protocluster.

\subsection{Ancillary Spectroscopic Data} \label{ancillary}

In the vicinity of the protocluster (150.12 $<$ R.A. (deg) $<$ 150.28, +1.92 $<$ decl. (deg) $<$ +2.08, 2.21 $<$ $z$ $<$ 2.25), we find 12 sources with spectroscopic redshift measurements from the zCOSMOS-deep survey (Lilly et al. in prep., also see \citealp{Lilly09}). We consider these as potential cluster members in addition to our observations. In Table \ref{table1}, we denote these extra sources by the label ``ancillary.'' 
  
\begin{figure*}
\centering
\includegraphics[width=7.0in]{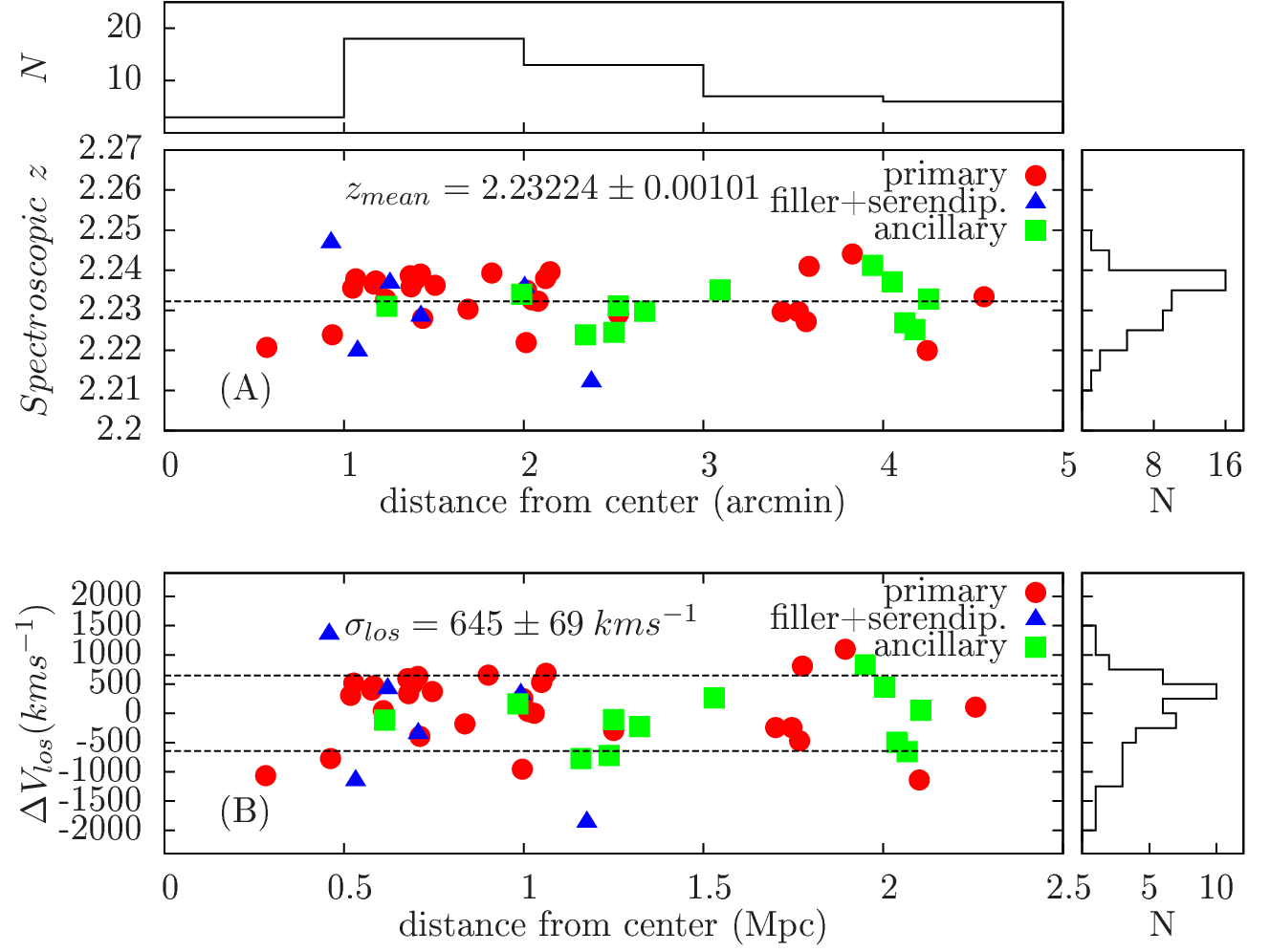}
\caption{(A) Redshift distribution of confirmed members (circles are primaries, triangles are fillers and serendipitous sources) as a function of projected distance (in arcmin) from the center of our protocluster CC2.2. $z_{mean}$ of the protocluster is shown with a black dashed line. (B) Line-of-sight velocity distribution with respect to the mean redshift as a function of projected distance (in Mpc) from the center. $\sigma_{los}$ boundaries for the member galaxies are shown with black dashed lines.}
\label{fig:z}
\end{figure*}

\section{Protocluster Characteristics} \label{property}

\subsection{Redshift and Velocity Dispersion} \label{velocity}

To select the protocluster members, we first determine the mean redshift and standard deviation of all unique galaxies (primary, filler, serendipitous, and ancillary). Sources that are within three standard deviations of the mean redshift are then used to determine the new mean redshift and standard deviation. We iteratively repeat this process until a final mean redshift ($z_{mean}$) and standard deviation ($\sigma_{z}$) is obtained. Only three galaxies (commented as ``field'' in Table \ref{table1}) do not pass the selection criterion. With the remaining 47 galaxies (35 from our observation and 12 from ancillary data), we estimate the mean redshift, line-of-sight dispersion in redshift space, and line-of-sight velocity dispersion ($\sigma_{los}$=c$\sigma_{z}$/(1+$z$) where c is the speed of light) as $z_{mean}$=2.23224 $\pm$ 0.00101, $\sigma_{z}$= 0.00696 $\pm$ 0.00074, and $\sigma_{los}$=645 $\pm$ 69 km s$^{-1}$, respectively. The uncertainties are estimated using the bootstrap method with 10,000 resamples. If we only rely on the primary sources (29 galaxies), we obtain $z_{mean}$(primary)=2.23321 $\pm$ 0.00113, $\sigma_{z}$(primary)=0.00615 $\pm$ 0.00073, and $\sigma_{los}$(primary)=570 $\pm$ 67 km s$^{-1}$, consistent with measurements using all the galaxies. 

To investigate the role of a small sample size on the results, following \cite{Yuan14}, we randomly select only 10 galaxies from our 47 members and recalculate the velocity dispersion. We estimate the new bootstrapped velocity dispersion as $\sigma_{los}(bootstrap)$=589 $\pm$ 149 km s$^{-1}$, consistent with what we found using the full sample, but with larger uncertainties. Figure \ref{fig:z} shows the redshift distribution, mean redshift, line-of-sight velocity distribution with respect to the mean redshift, and $\sigma_{los}$ boundaries for our member galaxies.  

\begin{figure*}
\centering
\hspace*{-0.70in}
\includegraphics[width=4.5in]{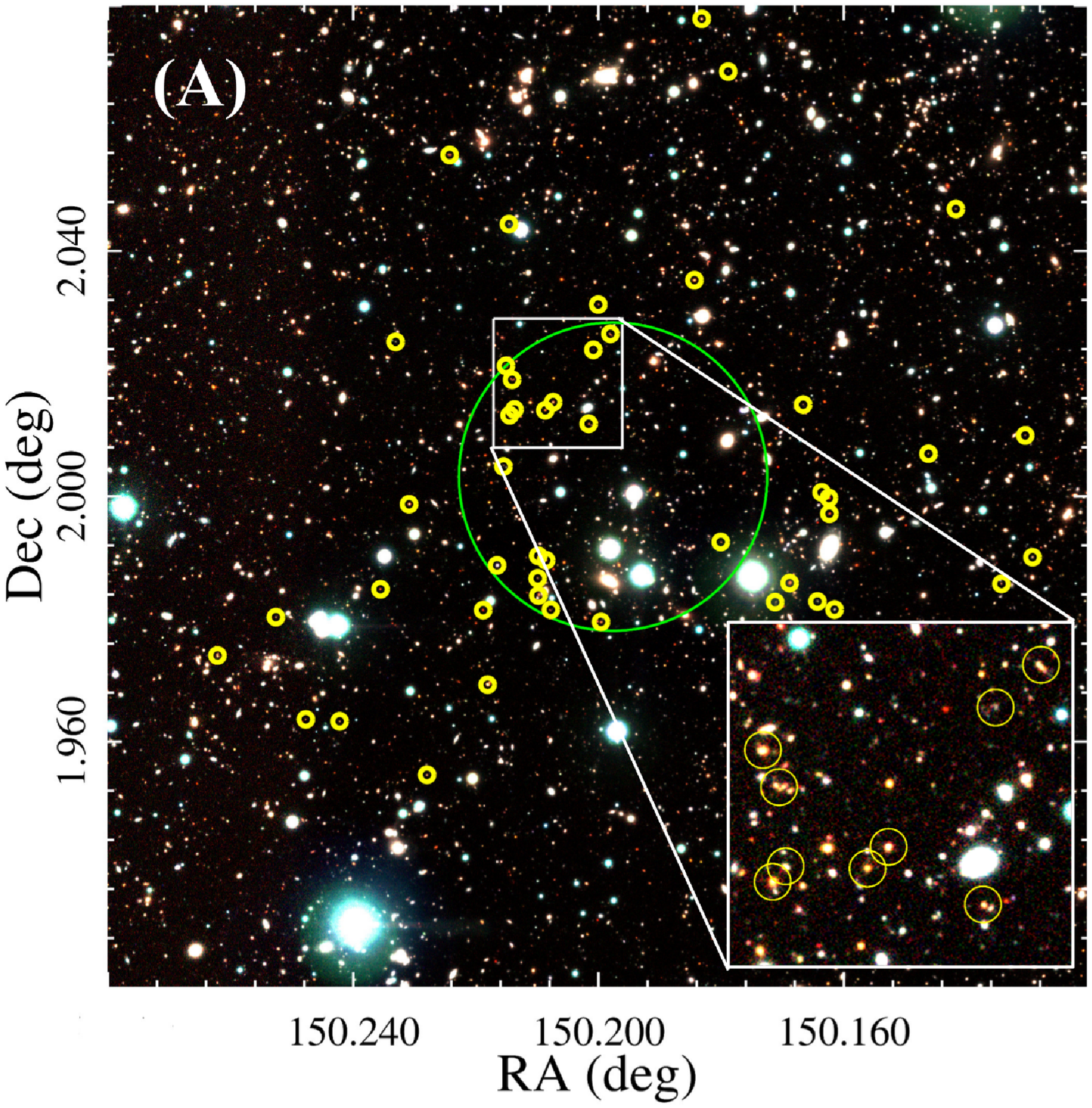}
\includegraphics[width=5.8in]{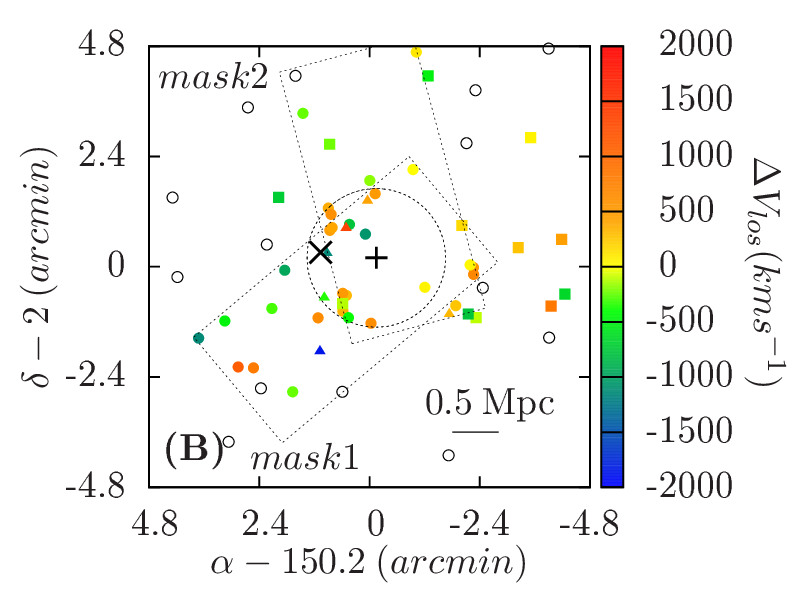}
\caption{(A) Three-color RGB image in the vicinity of our protocluster CC2.2. Yellow circles show the spatial distribution of the members. The green circle corresponds to the $R_{proj}$ of the protocluster. The red, green, and blue channels correspond to the UltraVISTA $K_{s}$, $J$, and $Y$ bands, respectively \citep{McCracken12}. (B) Spatial distribution of the protocluster members (circles are primaries, triangles are fillers and serendipitous sources, and squares are ancillary sources) color-coded by their line-of-sight velocities with respect to the mean redshift. The primary sources not observed (here in this paper or as ancillary) are shown with empty circles. The positions of the spectroscopic masks are shown with dashed rectangles. The plus sign shows the protocluster center. The dashed circle shows the estimated $R_{proj}$ of the protocluster. The multiplication sign shows the position of a candidate cluster (SACS-COSMOS-J100052+020018, Rettura et al. in prep.) seen as an overdensity of {\it Spitzer}-detected galaxies, reinforcing the reality of the structure.}
\label{fig:spatial}
\end{figure*}

\subsection{Spatial Distribution} \label{spatial}

We consider the centroid of the selected protocluster members as the protocluster center at R.A.=150.197509 (deg) and decl.=+2.003213 (deg). The centroid is defined as the arithmetic mean of the Cartesian unit vectors representing the protocluster members. For a 2D Gaussian distribution, $\sim$ 40\% of the weight of the distribution is within one standard deviation. Hence, we use the projected radius from the protocluster center that contains 40\% of the members as a proxy for the typical radius of the core of the protocluster and estimate it to be $R_{proj}$=0.75 $\pm$ 0.11 Mpc. Using only primary sources, we obtain R.A.(primary)= 150.208397 (deg), decl.(primary)=+2.000796 (deg), and $R_{proj}$(primary)=0.65 $\pm$ 0.13 Mpc. Fig \ref{fig:spatial} (A) shows the spatial distribution of the members. In Fig \ref{fig:spatial} (B), they are color-coded by the line-of-sight velocities relative to the mean redshift of the protocluster. We find that 51(87)\% of members are within 1(2) Mpc from the protocluster center.   

The match to the COSMOS2015 catalog shows that three of the members, mask1-1, mask1-15, and mask1-16 have {\it Chandra} X-ray detections \citep{Elvis09,Civano16,Marchesi16}. This comprises 6.8\%$\pm$3.7\% (6.9\%$\pm$4.9\%) of the members (primary members), a factor of $\sim$ 4 larger than the overall fraction of X-ray detected H$\alpha$ emitters in the HiZELS/COSMOS field at $z$=2.23 \citep{Calhau17}. All three have broad emission lines, indicative of their AGN nature and they are all Ly$\alpha$ emitters as well \citep{Matthee16,Sobral17}. The enhanced fraction of X-ray detected AGN in the protocluster relative to the field is in good agreement with \cite{Lehmer13}. mask1-16 also has a VLA 20 cm radio detection \citep{Schinnerer10}. These indicate that highly rare and active systems, such as extreme X-ray sources and radio galaxies trace dense environments at high-$z$, further supporting the dense nature of the protocluster. A detailed analysis of the AGN fraction will be presented in a following paper.

\subsection{Dynamical Mass} \label{mass}

One major difference between protoclusters and clusters, as discussed in, e.g., \cite{Diener15} and \cite{Wang16}, is that protoclusters are not yet fully virialized. Hence, for such nonvirialized systems, the velocity dispersion is mainly an indicator of the dynamical state of the system rather than the halo mass. Therefore, any estimation of the dynamical mass based on the velocity dispersion for nonvirialized systems should be considered as order-of-magnitude estimates and should be used with caution.
  
If we assume that the protocluster is virialized (see Section \ref{fate}) and $\sigma_{3d}$ and $R_{proj}$ are the total velocity dispersion and characteristic radius of the protocluster's core, then we can estimate its virial mass from the virial theorem as $M_{vir}$=$R_{proj}\sigma_{3d}^{2}$/G, where G is the gravitational constant. Assuming a spherical symmetry, $\sigma_{3d}^{2}$=3$\sigma_{los}^{2}$. Substituting $R_{proj}$ and $\sigma_{los}$ into the equation gives $M_{vir}$=(3$R_{proj}\sigma_{los}^{2}$/G)=(2.2 $\pm$ 0.6) $\times$ 10$^{14}$ $M_{\odot}$. With primary sources, we obtain $M_{vir}$(primary)=(1.5 $\pm$ 0.5)$\times$10$^{14}$ $M_{\odot}$. 

We can alternatively estimate the virial mass if we assume that the virial theorem applies to the protocluster and the halo of the protocluster is a spherical region within which the average density is 200$\rho_{c}(z)$, where $\rho_{c}(z)$ is the critical density of the universe at redshift of $z$ \citep{Navarro97}. Then, we can express the virial mass ($M_{200}$) of the protocluster in terms of its virial radius $r_{200}$ and the critical density as $M_{200}$=$\frac{4\pi}{3}r_{200}^{3}200\rho_{c}(z)$. The critical density can be expressed in terms of the Hubble parameter ($H(z)$) as $\rho_{c}(z)$=3$H^{2}(z)$/(8$\pi$G). Assuming a spherical symmetry combined with the virial theorem implies $r_{200}$=G$M_{200}$/(3$\sigma_{los}^{2}$). Therefore, we can express $r_{200}$ and $M_{200}$ as functions of $\sigma_{los}$ and $H(z)$ as $r_{200}$=$\sqrt{3}\sigma_{los}$/(10$H(z)$) and $M_{200}$=$(\sqrt{3}\sigma_{los})^{3}$/(10G$H(z)$) \citep{Carlberg97}. We estimate $r_{200}$=0.49 $\pm$ 0.05 Mpc and $M_{200}$=(1.4 $\pm$ 0.5)$\times$10$^{14}$ $M_{\odot}$.  Using only primary sources, $r_{200}$(primary)=0.43 $\pm$ 0.05 Mpc and $M_{200}$(primary)=(1.0 $\pm$ 0.3)$\times$10$^{14}$ $M_{\odot}$. These are in good agreement with $R_{proj}$ and $M_{vir}$ found above.

The {\it Spitzer} Archival Cluster Survey (SACS) is a comprehensive search for distant galaxy clusters in all {\it Spitzer}/IRAC extragalactic pointings available in the mission archive (Rettura et al. 2020 in prep.). Using the algorithm described in \cite{Rettura14}, high-redshift clusters are identified as overdensities in the mid-infrared data combined with shallow all-sky optical data. We find a match in their catalog (at a similar redshift), cluster SACS-COSMOS-J100052+020018, separated by only $\sim$ 1.2$^{\prime}$ from our protocluster. The position of their candidate is shown with a multiplication sign in Figure \ref{fig:spatial}. This provides further confirmation for the existence of the detected structure as Rettura et al. use a completely independent approach in finding high-$z$ protoclusters. Based on a relation calibrated in \citet{Rettura18} (see their Eq. 6), they use the {\it Spitzer} 4.5 $\mu$m richness of their clusters to infer their dynamical mass. This candidate cluster has an estimated mass, log($M_{500}/M_{\odot}$) $= 14.06 \pm 0.25$, consistent with our estimate based on the velocity dispersion.

Using simulated clusters, \cite{Munari13} suggest a scaling relation as $M_{200}/10^{15}M_{\odot}$=($\sigma_{1D}/A_{1D}$)$^{\alpha^{-1}}$/$h(z)$, where $A_{1D}$ and $\alpha$ are two parameters,  $\sigma_{1D}$ is the 1D velocity dispersion, and $h(z)=H(z)/H_{0}$. According to their Figure 3, $A_{1D}$ $\sim$ 1185 $\pm$ 30 km s$^{-1}$ and $\alpha$ $\sim$ 0.38 $\pm$ 0.01 at $z$=2 using galaxies as a tracer for the total mass of clusters. With this scaling relation, we obtain $M_{200}$(scaling)=(0.6 $\pm$ 0.2)$\times$10$^{14}$ $M_{\odot}$, consistent with $M_{200}$ we found before.  
   
\begin{table}
\begin{center}
\caption{Protocluster CC2.2 Characteristics} 
\begin{scriptsize}
\centering
\begin{tabular}{lccc}
\hline
\hline
\noalign{\smallskip}
Quantity & All Members & Primary Members Only\\
& & \\
\hline
\\
R.A. (deg)& 150.197509 & 150.208397\\
Decl. (deg) & +2.003213 & +2.000796\\
$z_{mean}$ & 2.23224$\pm$0.00101 & 2.23321$\pm$0.00113\\
$\sigma_{los}$(km s$^{-1}$) & 645$\pm$69 & 570$\pm$67\\
$R_{proj}$(Mpc) & 0.75$\pm$0.11 & 0.65$\pm$0.13\\
$M_{vir}$(10$^{14}$ $M_{\odot}$) & 2.2$\pm$0.6 & 1.5$\pm$0.5\\
$r_{200}$(Mpc) & 0.49$\pm$0.05 & 0.43$\pm$0.05\\
$M_{200}$(10$^{14}$ $M_{\odot}$) & 1.4$\pm$0.5 & 1.0$\pm$0.3\\
\\
\hline
\label{table2}
\end{tabular}
\end{scriptsize}
\end{center}
\end{table}
          
We note again that we have made a number of assumptions, such as virialization and the spherical symmetry in estimating the dynamical quantities. These assumptions may not be entirely correct, particularly for protoclusters at high redshift as they are likely still forming (see Section \ref{fate}). Therefore, these should be considered as order-of-magnitude estimates of the protocluster mass. In Table \ref{table2}, we summarize the protocluster CC2.2 characteristics using all the members and primary sources only.

\subsection{Protocluster's Fate} \label{fate}

Is the protocluster relaxed and fully virialized by the time of observation ($z$ $\sim$ 2.23)? The redshift distribution is not  symmetrically Gaussian (skewness=-0.5262, although the difference from a normal distribution is at $<$ 1.6$\sigma$ significance level) and the line-of-sight velocities with respect to the mean redshift are not fully symmetric (Figure \ref{fig:z}), indicating that the structure is still in the assembly process. As shown in Figure \ref{fig:map}, the presence of other potential overdensities and filamentary-like structures in the vicinity of the protocluster further suggests that the structure is likely not relaxed at $z$ $\sim$ 2.23 and still coalescing. 

We estimate the dynamical timescale ($\tau_{dyn}$) of the protocluster. The protocluster could be virialized at $z$ $\sim$ 2.23 if at least one dynamical timescale (in practice, a few) has elapsed since its formation. We estimate $\tau_{dyn}$ $\sim$ $r_{3d}$/$\sigma_{3d}$ where $r_{3d}$ is the characteristic radius of the protocluster and $\sigma_{3d}$ is its total velocity dispersion. If we assume $r_{3d}$ $\sim$ $R_{proj}$ and the spherical symmetry and use the estimated line-of-sight velocity dispersion and $R_{proj}$ from Section \ref{velocity}, we obtain $\tau_{dyn}$ $\sim$ 0.75 Mpc/($\sqrt{3}$ $\times$645 km s$^{-1}$) $\sim$ 0.6 Gyr. Therefore, if the protocluster was initially formed prior to $z$ $\sim$ 2.8, it would have had sufficient time to get virialized by the time of observation. Estimating the formation epoch of the protocluster is not straightforward. However, the average age of the stellar populations of its member galaxies, particularly the quiescent systems can place robust constraints on its formation time. By selection, quiescent galaxies are currently missing in our spectroscopic observation. However, future deep follow-up spectroscopic observations of potential passive galaxies in the protocluster can put stringent constraints on its formation epoch.  

Is the protocluster relaxed by now ($z$=0)? To answer this, we investigate the evolution of the protocluster overdensity in the linear regime of a spherical collapse model and compare it with the critical collapse threshold of $\delta_{c}$=1.69 \footnote{We note that this value of linearly extrapolated critical density enhancement is for an Einstein-de Sitter cosmology. However, it has been shown to have a weak dependence on cosmological models \citep{Percival05}.}.   
Within a redshift slice of $\Delta z$ $\sim$ 0.03 (width of the narrowband filter) and a projected 2 Mpc radius circle placed at the center of the protocluster, we find 35 H$\alpha$ emitters (the original sample from which the primary targets were selected for spectroscopy). The average number of H$\alpha$ emitters in the same volume is $\sim$ 4.6 (corrected for the effective area of the survey and the enhancement due to the overdensity). Therefore, using narrowband selected H$\alpha$ emitters, the galaxy number density enhancement is $\delta_{g}$=$\frac{35-4.6}{4.6}$=6.6.

Following \cite{Steidel05}, $\delta_{g}$ is related to the mass density enhancement ($\delta_{m}$) via 1+b$\delta_{m}$=C(1+$\delta_{g}$), where b is the clustering bias and C is a correction term due to the redshift space distortions and is calculated using C=1+$f$-$f$(1+$\delta_{m}$)$^{1/3}$, where $f$=$\Omega_{m}(z)^{0.6}$. Using $f$($z$=2.23)=0.96 and the clustering bias of b=2.4 for the H$\alpha$ emitters at $z$ $\sim$ 2.23 \citep{Geach12}, we obtain $\delta_{m}$($z$=2.23) $\sim$ 1.61. In a spherical collapse model \citep{Mo96}, this is related to a linear matter enhancement of $\delta_{L}$($z$=2.23) $\sim$ 0.73 and is expected to grow to $\delta_{L}$($z$=0) $\sim$ 1.9 by $z$=0. This exceeds the collapse threshold of $\delta_{c}$=1.69. Therefore, the protocluster is expected to fully collapse and virialize by now ($z$=0). In fact, the linear matter enhancement reaches the collapse threshold at $z$ $\sim$ 0.1, indicating that the protocluster should have been virialized since the past $\sim$ 1.0-1.5 Gyr. The collapse threshold at any redshift is approximated as $\delta_{c}(z)$ $\simeq$ 1.69$D$($z$=0)/$D(z)$, where $D(z)$ is the linear growth function \citep{Percival05}. At the redshift of the protocluster, $\delta_{c}(z=2.23)$ $\sim$ 4.3. This is larger than $\delta_{L}$($z$=2.23), further indicating that the structure is likely not virialized at $z$=2.23.   

We estimate the virialized mass of the protocluster at present $M_{dyn}$($z$=0) through $M_{dyn}$($z$=0) = $\rho_{m}(V_{obs}/C)(1+\delta_{m})$, where $\rho_{m}$ is the mean comoving density, $V_{obs}$ is the observed comoving volume of the structure, and C is the correction term introduced above \citep{Steidel98}. The H$\alpha$ emitter candidates are dominated by those selected in the UKIRT/WFCAM narrowband $K$ filter \citep{Sobral13}. Assuming a tophat shape for the filter corresponds to a redshift width of $\Delta z$ $\sim$ 0.032 centered at $z$ $\sim$ 2.23. The corresponding comoving radial width ($\Delta \chi$) is then $\sim$ 42 Mpc. This leads to the comoving $V_{obs}$ $\sim$ 5500 Mpc$^{3}$ for a cylinder of width $\Delta \chi$ and a projected physical radius of 2 Mpc at $z$ $\sim$ 2.23. Given $\delta_{m}$($z$=2.23) $\sim$ 1.61 and C=0.64, we estimate $M_{dyn}$($z$=0) $\sim$ 9.2$\times$10$^{14}$ $M_{\odot}$. Therefore, the protocluster is likely the progenitor of a Coma-type cluster at $z$=0. Simulations of \cite{Chiang13} show that at $z$ $>$ 2, the progenitors of a Coma-type cluster traced by SFR $>$ 1 $M_{\odot}$yr$^{-1}$ galaxies are expected to have a galaxy density enhancement of $\delta_{g}$ $\sim$ 5.5$^{1.5}_{-0.8}$ probed over 15$^{3}$ $\sim$ 3500 Mpc$^{3}$ comoving volumes. These values are in rough agreement with our measurements, indicating that our protocluster is expected to evolve into a $\sim$ 10$^{15}$ $M_{\odot}$ Coma-type cluster at $z$=0.

The comoving volume associated with H$\alpha$ emitter candidates in the HiZELS/COSMOS field is $\sim$ 5.48 $\times$ 10$^{5}$ Mpc$^{3}$ \citep{Sobral13}. Given the detection of one protocluster in this volume, we estimate a comoving space and mass density of $\sim$ 1.8 $\times$ 10$^{-6}$ Mpc$^{-3}$ and $\sim$ $(1.8-3.6) \times$10$^{8}$ $M_{\odot}$ Mpc$^{-3}$ for a $M_{dyn}$ $\sim$ $(1-2) \times$10$^{14}$ $M_{\odot}$ protocluster at $z$ $\sim$ 2. However, we note that Poisson uncertainties are as large as the reported values. With the Poisson uncertainty, the space density of the protocluster is $\lesssim$ 3.6 $\times$ 10$^{-6}$ Mpc$^{-3}$. The halo mass function of \cite{Bocquet16} predicts a space density of $\sim$ 1-2 $\times$ 10$^{-7}$ Mpc$^{-3}$ for a $M_{200}$ $\sim$ 10$^{14}$ $M_{\odot}$ halo at $z$=2, a factor of $\sim$ 10 smaller than our estimate, but consistent with it given the large Poisson uncertainty in our measurement. Moreover, for a sample of similarly selected H$\alpha$ emitters in the UDS \citep{Sobral13} and Bo{\"o}tes \citep{Matthee17} fields at $z$ $\sim$ 2.2 with comoving volumes of $\sim$ 2.24 $\times$ 10$^{5}$ Mpc$^{3}$ and $\sim$ 2.7 $\times$ 10$^{5}$ Mpc$^{3}$, respectively, no overdensity of H$\alpha$ emitters is found. This increases the effective volume and subsequently decreases the space density of our protocluster, making our measurement more consistent with the halo mass function predictions.     

\section{Comparison} \label{comparison}

We compare the present-day mass of $z$ $\gtrsim$ 1.5 protoclusters compilation from \cite{Overzier16} with that of our protocluster. The median protocluster in the compilation has a present-day mass of log($M(z=0)/M_{\odot}$)=14.6. This makes our protocluster one of the most massive systems with a $z$=0 mass comparable to some remarkable high-$z$ protoclusters with $M$($z$=0) $\gtrsim$ 10$^{15}$ $M_{\odot}$ \citep{Venemans02,Cucciati14,Lee14,Lemaux14,Diener15,Badescu17,Oteo18,Chanchai19}. 

Protoclusters at $z$ $\sim$ 2-3 in the \cite{Overzier16} compilation have galaxy overdensities in the range $\delta_{g}$ $\approx$ 1.5-16, with those that used H$\alpha$ emitters as the tracer of overdensity have $\delta_{g}$ $\approx$ 4-16. We note that these values are measured differently with different selection functions and volumes probed. Therefore, these should not be directly compared with one another and our work. Nevertheless, they show that our protocluster overdensity of $\delta_{g}$ $\sim$ 7 is typical of high-$z$ protoclusters and they are in broad agreement with simulations of \cite{Chiang13}. 

\cite{Cucciati18} recently identified a super-protocluster in formation in the COSMOS field at $z$ $\sim$ 2.45, dubbed ``Hyperion,'' containing at least seven density peaks with masses in the range $\sim$ (0.1-2.7)$\times$10$^{14}$ $M_{\odot}$. Hyperion is extended over a comoving volume of $\sim$ 60$\times$60$\times$150 Mpc$^{3}$ and has an estimated total mass of $\sim$ 4.8$\times$10$^{15}$ $M_{\odot}$. Could the extended LSS shown in Figure \ref{fig:map} (A) be a super-protocluster similar to ``Hyperion''? The comoving radial distance between the northern cluster at $z$ $\sim$ 2.1 and our protocluster in the south is $\sim$ 180 Mpc. If the extended structure (including the central overdensity shown with a question mark and other potential surrounding overdensities) is confirmed to be a multicomponent super-protocluster, it would have a comoving volume of $\sim$ 40$\times$40$\times$180 Mpc$^{3}$, making it comparable to Hyperion. Follow-up spectroscopic observations could further reveal the nature of this structure.             
        
\section{Summary} \label{sum}

We report the spectroscopic confirmation of a new protocluster in the COSMOS field at $z$=2.23224, dubbed CC2.2, using Keck/MOSFIRE observations in combination with ancillary data from zCOSMOS-deep spectroscopic survey. With 47 confirmed members (35 from our MOSFIRE observations and 12 from ancillary data), we estimate a line-of-sight velocity dispersion and a total mass of $\sigma_{los}$=645 $\pm$ 69 km s$^{-1}$ and $M_{dyn}$ $\sim$ $(1-2) \times$10$^{14}$ $M_{\odot}$ for the protocluster, respectively. The structure is likely not fully virialized at $z$ $\sim$ 2.23 but is expected to collapse to a Coma-type cluster with $M_{dyn}$($z$=0) $\sim$ 9.2$\times$10$^{14}$ $M_{\odot}$ at $z$=0.  

With the high-quality data obtained, in forthcoming papers, we will investigate the role of early environments on the SFR (H$\alpha$ or H$\beta$), nebular extinction (H$\beta$ and H$\alpha$), gas-phase metallicity ([N{\sc ii}]$\lambda$6549, [N{\sc ii}]$\lambda$6583, and H$\alpha$), electron density ([S{\sc ii}]$\lambda\lambda$6717,6731 doublet), source of ionization (BPT diagram), ionization state of the gas ([O{\sc iii}]$\lambda$4959, [O{\sc iii}]$\lambda$5007, and H$\beta$), mergers, dynamics, and AGN fraction relative to galaxies in the field. Moreover, follow-up spectroscopy can further reveal the potential multicomponent nature of the structure shown in Figure \ref{fig:map}.
     
\section*{acknowledgements}

We are thankful to the anonymous referee for their useful comments and suggestions that improved the quality of this paper. B.D. acknowledges financial support from NASA through the Astrophysics Data Analysis Program (ADAP), grant number NNX12AE20G, and the National Science Foundation, grant number 1716907. B.D. is thankful to Andreas Faisst, Laura Danly, and Matthew Burlando for their companionship during the observing run. B.D. is grateful to the COSMOS team for their useful comments during the team meeting in New York City 2019 May 14-17. B.D. wishes to thank Dongdong Shi for bringing an issue in an earlier version to our attention. A.R. research was made possible by {\it Friends of W. M. Keck Observatory} who philanthropically support the Keck Science Collaborative (KSC) fund. The observations presented herein were obtained at the W. M. Keck Observatory (program C236, PI Scoville), which is operated as a scientific partnership among the California Institute of Technology, the University of California, and the National Aeronautics and Space Administration. The Observatory was made possible by the generous financial support of the W. M. Keck Foundation. The authors would like to recognize and acknowledge the very prominent cultural role and reverence that the summit of Maunakea has always had within the indigenous Hawaiian community. We are fortunate to have the opportunity to perform observations from this mountain.

\bibliographystyle{aasjournal} 
\bibliography{references}

\end{document}